%% file: dyadic.tex
\documentclass[12pt]{iopart}
\usepackage{epsfig}
\usepackage{graphicx}
\usepackage{dcolumn}
\usepackage{bm}
\usepackage{longtable}
\usepackage[countmax]{subfloat}
\usepackage{epstopdf}

\newcommand{\kvar}{{\bra k^2 \ket}}

\input{commands}

\begin{document}

\title{The two-star model: exact solution in the sparse regime 
and condensation transition}

\author{A Annibale$^\dag$, OT Courtney}
\address{$\dag$ ~ Department of Mathematics, King's College London, The Strand,
London WC2R 2LS, UK}

\date{\today}

\begin{abstract}
The $2$-star model is the simplest exponential random graph model that 
displays complex behavior, such as degeneracy and phase transition. Despite its 
importance, this model has been solved only 
in the regime of dense connectivity.
In this work we solve the model in the 
finite connectivity regime, far more prevalent in real world networks.
We show that the model undergoes a condensation transition from a liquid to a condensate phase 
along the critical line corresponding, in the ensemble parameters space, to the Erd\"os-R\'enyi graphs.
In the fluid phase the model can produce graphs with a narrow degree statistics, ranging from regular to Erd\"os-R\'enyi graphs, 
while in the condensed phase, the ``excess'' degree heterogeneity condenses 
on a 
single site with degree $\sim\sqrt{N}$.
This shows the unsuitability of the two-star model, in its standard definition, to produce arbitrary finitely connected graphs with degree 
heterogeneity higher than Erd\"os-R\'enyi graphs and suggests that non-pathological variants of this model may be attained by 
softening the global constraint on the two-stars, while keeping the number 
of links hardly constrained.
\end{abstract}

\section{Introduction}

Exponential random graph models (ERGM) 
are ensembles of random graphs where each graph configuration 
$\bc$ 
appears with a probability $p(\bc)\propto e^{-H(\bc)}$
given by the Gibbs-Boltzmann distribution, where $H(\bc)$ is the graph 
Hamiltonian, enclosing several properties of the 
networks in the ensemble. 
First introduced in the $1980s$ by Holland 
and Leinhardt \cite{HolLei81}, and further developed by Frank and Strauss 
\cite{FraStr86} and in several later studies 
\cite{StraussIkeda90,WassermanPattison96,SanilBanksCarley95,BanksCarley96,
Snijders01}, ERGM soon became popular in social network analysis.
Computational 
tools to analyze and simulate networks based on ERGM 
are largely available on the web, as the ERGM and SIENA packages, and  
several paradigmatic models of random graphs can be written in the exponential 
form, for suitable choices of the graph Hamiltonian, including the Erd\"os-R\'enyii (ER) graph ensemble \cite{ErdRen59}
and the ensemble of graphs with soft-constrained degree sequence 
\cite{BenCan78,Bol80,MolRee95}.
However, ERGM have known
drawbacks which limit their practical use as proxies or null models for 
real networks. 
In particular, they may display degeneracy behaviour 
and may fail to produce graphs with properties within certain ranges, 
which are nevertheless observed in nature. 

A useful model to understand degeneracy behavior and limitations of ERGM,
is the two-star model, which is simple enough to be amenable of exact solutions,
while exhibiting the interesting bahavior of more complex 
ERGM.
A well-known solution for the two-star model has been developed, 
in the limit of large system size $N$,
by Park and Newmann \cite{ParkNew04}, 
within a mean-field approach and expansions around it. However, this approach requires that 
the connectivity of the 
graph is $\order{(N)}$, a condition which is hardly met by real 
world networks, like social and biological networks, where 
the connectivity is $\order{(N^0)}$. 
In this work we solve the two-star model exactly, in the finite 
connectivity regime, 
for large system size, by using path integrals.
We check our calculations against Monte-Carlo simulations and 
compare our results with those predicted by 
mean-field theories. Results show that the $2$-star model undergoes a condensation transition from a liquid to 
a condensate state, along the critical line corresponding, in the ensemble paramters space, to the Erd\"os-R\'enyi graphs. 
In the liquid phase the model can only 
produce graphs within a narrow range of degree statistics, namely 
between regular and Erd\"os-R\'enyi (ER), whereas in the condensed phase a condensate of size $\sim\sqrt{N}$ emerges, residing on a single site.

The paper is structured as follows: in Sec. \ref{sec:model} we 
define the model and review the mean-field solution, in Sec. \ref{sec:exact} we solve the two-star model 
in the finite connectivity regime and
compare our results with the mean-field predictions and 
with Monte-Carlo simulations. In Sec. \ref{sec:condensation} we show that the phase transition 
displayed by the system in the finite connectivity regime is a condensation, related to large fluctuations in the 
sums of random variables. 
Finally, we summarise our conclusions in Sec.
\ref{sec:conclusions}.

\section{The model and the mean-field solution}
\label{sec:model}
The $2$-star model is a prototypical example of ERGM. 
In this section, we give a brief review of ERGM, the $2$-star model and its mean-field solution.

\subsection{Background: the ERGM}
\label{sec:erg}
For simple and undirected graphs of $N$ nodes, ERGM 
are graph ensembles where    
each graph is defined by an adjacency matrix $\bc$, 
with elements $\cij\in\{0,1\} ~\forall~i,j=1,\ldots,N$, 
$c_{ii}=0~\forall~i$, $\cij=\cji~\forall~i<j$. 
In ERGM, each graph $\bc$ appears with a probability given by the Gibbs-Botzmann distribution
\be
p(\bc)=\frac{1}{Z}e^{-H(\bc)}
\label{eq:Gibbs}
\ee 
with Hamiltonian $H(\bc)$ and partition function $Z=\sum_\bc e^{-H(\bc)}$.
The Hamiltonian is given by 
\be
\label{eq:Hamiltonian}
H(\bc)=-\sum_{\mu=1}^K \lambda_\mu \Omega_\mu (\bc)
\ee
where $\bOmega(\bc)=(\Omega_1(\bc),\ldots,\Omega_K(\bc))$ is a set 
of observables of the graph $\bc$ for which one has 
statistical estimates $\bOmega=(\Omega_1,\ldots,\Omega_K)$, so that 
\be
\bra \Omega_\mu(\bc)\ket=\Omega_\mu\quad\quad \forall~\mu=1\ldots,K
\label{eq:constraints}
\ee
with $\bra \Omega_\mu(\bc)\ket=\sum_\bc p(\bc)\Omega_\mu(\bc)$ and $\blambda=(\lambda_1,\ldots,\lambda_K)$ are ensembles parameters also 
called ``conjugated'' 
observables, which have to be calculated
from the equations for the constraints (\ref{eq:constraints}). 
The distribution (\ref{eq:Gibbs}) with 
Hamiltonian (\ref{eq:Hamiltonian}) maximises the Shannon entropy 
\be
{\mathcal S}(\bOmega)=-\sum_\bc p(\bc)\ln p(\bc)
\ee
subject to the constraints (\ref{eq:constraints}) and the 
normalization $\sum_\bc p(\bc)=1$.
Hence, ERGM are maximum entropy ensembles 
conditioned on the imposed constraints.

For all but the simplest ERGM, exact solutions for the ensemble 
parameters are not available, so one 
has normally to take recourse to numerical methods. 
The latter have been 
the subject of intense investigation over the last few decades 
and range from pseudo-likelihood 
estimation \cite{Besag75,StraussIkeda90,AndWasCro99,PattisonRobins02,WassermanPattison96,WassermanRobins05} to Markov Chain  
Monte Carlo Maximum Likelihood techniques \cite{Snijders02,Handcock03,SnijdersRecent,WassermanRobins05}, 
to Bayesian inference \cite{AustadFriel10,CaimoFriel11,CaimoFriel13}.

Once the values of the parameters $\blambda$ are available, one can 
use the resulting probability distribution $p(\bc)$ to estimate the value 
of observables for which no estimate is available.
We note that expectation values for the primary network observables 
$\{\Omega_\mu\}$ are given by 
\bea
\bra \Omega_\mu(\bc)\ket=\frac{1}{Z(\blambda)}
\sum_\bc \Omega(\bc) e^{\sum_\mu \lambda_\mu \Omega_\mu(\bc)}
=
-\frac{\partial F(\blambda)}{\partial \lambda_\mu}
\eea
where $F(\blambda)=-\ln Z(\blambda)$ is the free-energy, and their fluctuations 
are given by the second derivative of the free-energy $\bra \Omega_\mu^2\ket-\bra \Omega_\mu\ket^2
=\partial^2 F/\partial \lambda_\mu^2$
which, in equilibrium, equates the so-called susceptibility, defined as 
$
\chi_{\mu}=\partial \bra \Omega_\mu\ket/\partial \lambda_\mu,
$
and measuring the deviation in $\bra \Omega_\mu\ket$ when a change 
in the external variable $\lambda_\mu$ is applied. 
These relations suggest that in general we are able to calculate the 
ensemble parameters from the equations for the constraints, if we 
can calculate the free energy 
of the ERGM. This is straightforward only for 
Hamiltonians which are linear in the adjacency matrix, where the 
graph distribution factorizes over the links $p(\bc)=\prod_{i<j}p(c_{ij})$. 
For non-linear Hamiltonians, like the two-star model and Strauss model 
\cite{StraussIkeda90}, 
solutions have been found within mean-field approximations and 
expansions about it \cite{ParkNew04,ParkNewman05}. 

\subsection{The $2$-star model}
\label{sec:twostar}
In the two-star model, one chooses, as graph observables, 
the number of links 
\be
L(\bc)=\half \sum_{i\neq j}\cij
\label{eq:L}
\ee
and the number of two-stars (i.e. paths of length two)
\be
S(\bc)=\frac{1}{2}\sum_{i\neq j\neq k(\neq i)}\cij\cjk
\label{eq:S}
\ee
and assumes that the expectations $L=\bra L(\bc)\ket$ and $S=\bra S(\bc)\ket$ are known.
This leads to the non-linear network Hamiltonian
\bea
H(\bc)&=&-\theta_1 L(\bc)-\theta_2 S(\bc)
=-\sum_{i<j}\cij(\theta_1+\theta_2\sum_{k\neq i,j}\cjk)
\nonumber\\
&=&-\sum_{i<j}\cij\left[\left(\theta_1-\theta_2\right)+\theta_2\sum_k \cjk\right]
\nonumber\\
&=&-\sum_{i<j}\cij\left(2\alpha+2\beta\sum_k \cjk\right)
\label{eq:HMF}
\eea
where we set $\alpha=(\theta_1-\theta_2)/2$ and $\beta=\theta_2/2$.
Alternatively, we can define the local degrees of graph $\bc$ as 
$k_i(\bc)=\sum_j \cij~\forall ~i$, and write 
the observables (\ref{eq:L}) and (\ref{eq:S}) as 
\bea
L(\bc)&=&\half \sum_i k_i(\bc)
\nonumber\\
S(\bc)&=&\frac{1}{2}(\sum_j k_j^2(\bc)-\sum_j k_j(\bc))
\eea
so that the Hamiltonian reads
\bea
H(\bc)
&=&-\alpha\sum_i k_i(\bc)-\beta\sum_j k_j^2(\bc).
\eea
At this point it is useful to define the 
average connectivity $\bark(\bc)=N^{-1}\sum_i k_i(\bc)$ and  
the expected average connectivity in the ensemble
$\kav=\bra \bark(\bc)\ket$, with 
$\bra \cdot\ket$ denoting the average over the
ensemble probability $p(\bc)$.
In the two-star model, we have $L=N\kav/2$ and $S=N(\kvar-\kav)/2$,
hence we constrain $\kav$ and $\bra k^2 \ket$.

The two-star model has so far been solved for dense graphs, with
average connectivity $\kav=\order{(N)}$,  
for which 
mean-field approaches are exact in the thermodynamic limit.
The mean-field solution (see Section \ref{sec:PN}) shows that for large $N$, and $\beta N=2J$,
the two-star model undergoes, for any $J \geq 1$, 
a first-order phase transition,  
between a phase of low connectivity and one of 
high connectivity, separated by the critical line $J=-\alpha$ in the space of ensemble 
parameters. 
The critical line terminates at the critical point $J=1$, 
where a second order phase transition takes place from the symmetry-broken 
state with two phases, the one with high and the one with low connectivity respectively, to a symmetric state
\cite{ParkNew04,BizPacGra12}. In particular, for $J\geq 1$ 
the link density is discontinuous meaning that the model 
fails to produce intermediate connectivities by suitably tuning the ensemble parameters. Along the critical line 
$J=-\alpha$ (and for $J\geq 1$) one has degeneracy, meaning that for the same ensemble parameters the model produces 
either a sparse or a dense graph.

However, it is not clear a priori how well the mean-field scenario applies to the finite connectivity regime, 
where Gaussian fluctuations (around the mean-field solution) become dominant rather than a small 
perturbation around the leading order statistics. 
In addition, we are interested in establishing whether in the phase at low connectivity the 
model can produce arbitrary densities of stars for any given (finite) connectivity, by 
choosing suitably the ensemble parameters. More in general, we aim to establish whether the $2$-star model may serve 
as a plausible null-model for finitely connected networks, which are far more prevalent than dense networks in the real world.
We answer these questions precisely in Section \ref{sec:exact} by solving the two-star model exactly, in the limit $N\to \infty$, 
for finite connectivity $\kav=\order{(N^0)}$.

\subsection{The mean-field approach}
\label{sec:PN}
In this section we briefly illustrate the mean-field approach, 
that is exact, in the thermodynamic limit, for dense graphs \cite{ParNew04b}, but it is expected to give inaccurate results 
for finite connectivity.
By analogy with spin models, we can regard the expression in the brackets of (\ref{eq:HMF})
as the local field acting on the link $\cij$. The mean-field approximation replaces the local fields with their ensemble averages, i.e. $\cjk\to\bra \cjk\ket$. 
Doing so, all edges in the model become equivalent and the average probability to observe a link is the 
same for all links $p=\bra \cjk\ket~\forall~j,k$. Hence, the Hamiltonian becomes 
\be
H(\bc)=-\lambda \sum_{i<j}\cij
\label{eq:Ham}
\ee
where 
\be
\lambda=2\alpha+2\beta(N-1)p\simeq 2\alpha + 2\beta N p
\label{eq:lambda}
\ee
is now a function of the unknown probability $p$ and the last approximation holds for $N\gg 1$.
The Hamiltonian (\ref{eq:Ham}) leads to the partition function
\be
Z=\sum_\bc e^{\lambda \sum_{i<j}c_{ij}}=\prod_{i<j}\sum_{c_{ij}}e^{\lambda c_{ij}}=\prod_{i<j}(1+e^\lambda)=(1+e^\lambda)^{N(N-1)/2}
\ee
which immediately gives the free energy
\be
F=-\ln Z=-\frac{N(N-1)}{2}\ln(1+e^\lambda)
\label{eq:MF_f}
\ee
that can be used to write the equation for the constraint 
\be
L=-\frac{\partial F}{\partial \lambda}=\frac{N(N-1)}{2}\frac{e^\lambda}{1+e^\lambda}.
\label{eq:Lav}
\ee
If links are all drawn randomly and independently with probability $p$, we have 
\be
L=\frac{N(N-1)}{2}p
\ee
hence we get from (\ref{eq:Lav})
\be
p=\frac{e^\lambda}{1+e^\lambda}=\half \left(1+\tanh \frac{\lambda}{2}\right)
\label{eq:MF_p}
\ee
where in the last equality we used the identity $e^x/2\cosh x=(1+\tanh x)/2$.
Now, however, $\lambda$ is a function of $p$, so the above gives a self-consistency equation for $p$
\be
p=\half \left[\tanh(\beta N p+\alpha)+1\right]
\label{eq:selfp}
\ee
This equation is identical to the one found by Park and Newman 
\cite{ParNew04b}
by solving the model exactly in the dense regime, i.e. for $p=\order{(1)}$, and by setting $\beta N=2J$
\begin{equation}
p=\frac{1}{2}\Big[\tanh\big(2Jp + \alpha\big) + 1 \Big]
\label{eq:PN}
\end{equation}
As noted in \cite{ParNew04b}, 
a unique solution to (\ref{eq:PN}) exists only for $J<1$. 
For $J\geq 1$ and $\alpha$ sufficiently close to $-J$ there 
are three solutions, with only the outer two being stable, leading to a degeneracy in the solution.
Expansions around the mean-field solution and perturbation theories \cite{ParkNew04,ParkNewman05} give for the first two moments of the degree distribution
\bea
\kav&=& Np+\frac{4Jp(1-p)(1-2p)}{[1-8Jp(1-p)[1-4Jp(1-p)]}
\label{eq:MF_kav}\\
\bra k^2\ket &=&
N^2p^2+\frac{Np(1-p)(1-8Jp^2)}{[1-8Jp(1-p)[1-4Jp(1-p)]}
\label{eq:MF_kvar}
\eea
where the second terms on the RHS, originate from the Gaussian fluctuations about the mean-field solution. These are subleading for large $N$, 
in the dense regime 
where $p=\order{(1)}$ and $J=\order{(1)}$.
However, in the finite connectivity regime where $p=\order{(N^{-1})}$ and $\beta=\order{(1)}$ (see \ref{app:graphical}), Gaussian 
fluctuations are no longer small fluctuations about the leading order statistics.
Upon setting $p=c/N$ and sending $N\to \infty$ at constant $\beta$ and $c$, we get 
\bea
\kav&=& c\left[
1+\frac{\beta}{(1-2\beta c)(1-\beta c)}
\right]
\label{eq:MF_kav}
\\
\bra k^2\ket &=&
c\left[c+\frac{1}{(1-2\beta c)(1-\beta c)}
\right]
\label{eq:MF_kvar}
\eea
showing that the mean-field approximation becomes inexact in the finite 
connectivity regime. 
For a full discussion of mean-field predictions in the finite connectivity regime see \ref{app:graphical}.

\subsection{Upper and lower bounds on the total number of stars in the finite connectivity regime}
\label{sec:minmax}
Before solving the model in the finite connectivity regime, it is useful to derive expressions for the upper and lower physical bounds 
on the number of stars that the model can exhibit at a given finite connectivity.
The expected number of 2-stars is
$S=N(\bra k^2\ket-\kav)/2$.
For a 
fixed number of edges $L$, the total number of stars is minimised by minimising the second moments of the degree distribution 
while keeping the first moment fixed, i.e. 
by making the degree distribution regular, so that $\bra k^2\ket=\kav^2$ attains its physical 
minimum. This corresponds to the minimum star density  
\be
\frac{S_{\rm min}}{N} =\half \kav(\kav-1)
\ee
The total number of stars is maximised by maximising the 
second moment while keeping the first moment fixed, resulting in a small number of 
vertices having very large degrees while all the others have low degrees. To see this, 
we consider the following iterative process. We pick two vertices $i$ and $j$ and assume that $k_i\geq k_j$. If $j$ has a neighbour which is
not already connected to $i$ then this edge is rewired to increase $k_i$ by $1$ and decrease $k_j$ by 1. 
This step is repeated until in every pair, the vertex with the lesser degree has no more 'spare' edges to rewire to the other vertex. 
In the case where $L\leq N-1$ this always ends with a graph where there is one vertex with degree $L$, $L$ vertices connected to it with degree 1, and any left over vertices having degree 0 (because any other configuration would contain 'spare' edges). If we increase $L$ just beyond $N$, we can no longer increase the stars by rewiring to the original hub as that hub is already connected to every other vertex. This causes a new hub to form to take up the extra edges. This process of hubs being born and increasing in size until they have degree $N$ keeps going as $L$ is increased until the graph is complete.

In this work we consider sparse graphs, where the average connectivity is $\order{(1)}$. For this case the total number of edges $L$ is $O(N)$ which means that the minimum 
value that the number of stars can take is $O(N)$. On the other hand, the maximum star configuration has an $O(1)$ number of vertices with degree $N$, and an $O(N)$ number of vertices with degree $O(1)$. The vertices with degree $N$ give 
each a contribution to the number of stars $N(N-1)/2$ showing that the maximum density of stars
in a finitely connected graph is
\be
\frac{S_{\rm max}}{N} \sim \order{(N)}.
\ee
Expansions about mean-field solutions (\ref{eq:MF_kav}), (\ref{eq:MF_kvar}) show that both moments $\kav$, $\bra k^2\ket$ diverge
at the same parameter values, suggesting that it is impossible to achieve the maximum expected star density for any finite value of connectivity. 
However, higher order (non-Gaussian) fluctuations may become dominant in the finite connectivity regime and the expansions (\ref{eq:MF_kav}) and (\ref{eq:MF_kvar}) may get very inaccurate. 
We will test their accuracy against MCMC simulations and results of the exact calculation in the next section.


\section{Exact solution in the finite connectivity regime}
\label{sec:exact}
In this section, we derive an exact solution for the 
two-star model in the finite connectivity regime, where mean-field 
solutions are expected to become inexact. First off, we rewrite the 
graph Hamiltonian by performing the sum over $k$ in (\ref{eq:HMF}) 
and using symmetry of $c_{\ij}=c_{ji}$
\be
H(\bc)=-2\alpha\sum_{i<j}\cij-\beta \sum_{i<j}\cij(k_i(\bc)+k_j(\bc)).
\label{eq:H}
\ee
Hence, we introduce the partition function 
\bea
Z&=&\sum_\bc e^{2\alpha\sum_{i<j}\cij+\beta\sum_{i<j}(k_i(\bc)+k_j(\bc))c_{ij}}
\nonumber\\
&=&\sum_\bk\sum_\bc \delta_{\bk,\bk(\bc)}e^{\sum_{i<j}\cij(2\alpha+\beta(k_i+k_j))}
\nonumber\\
&=&\sum_\bk \int_{-\pi}^\pi \dO \sum_\bc e^{\sum_{i<j}\cij(2\alpha+\beta(k_i+k_j)-i(\Oi+\Oj))}
\nonumber\\
&=&\sum_\bk \int_{-\pi}^\pi \dO 
\prod_{i<j}(1+e^{2\alpha+\beta(k_i+k_j)-i(\Oi+\Oj)})
\nonumber\\
\label{eq:partition}
\eea
where $\delta_{\bk,\bk(\bc)}=\prod_i \delta_{k_i,k_i(\bc)}$ with $\delta_{x,y}$ being the Kronecher delta, taking value $1$ for $x=y$ and $0$ otherwise,
and the Fourier representation of the Kronecher delta has been used
\be
\delta_{x,y}=\int_{-\pi}^\pi \frac{d\Omega}{2\pi}e^{i\Omega(x-y)}.
\ee
Next, we focus on the normalised logarithm of (\ref{eq:partition}), giving the 
free energy density $f=-N^{-1}\log Z$, which can be calculated exactly, for large $N$, in the finite connectivity 
regime $\bra \cij\ket =\order{(N^{-1})}~\forall~i,j$, by using path integrals. 

As a first step, 
we consider the likelihood $\bra \cij\ket$ for two nodes $i,j$ to be connected. This follows from the estimate of the average number of links $L=\bra L(\bc)\ket=\sum_{i<j}\bra \cij\ket$. Using
\bea
p(\bc)&=&\frac{1}{Z}\sum_\bk \int_{-\pi}^\pi \dO 
\prod_{i<j} \left[e^{2\alpha+\beta(k_i+k_j)-i(\Oi+\Oj)} \delta_{\cij,1}+\delta_{\cij,0}\right]
\nonumber\\
\eea
we have
\bea
L&=&\sum_{i<j}\bra \cij\ket =\frac{1}{Z}\sum_{i<j}
\sum_\bk \int_{-\pi}^\pi \dO 
\frac{e^{2\alpha+\beta(k_i+k_j)-i(\Oi+\Oj)}}{1+e^{2\alpha+\beta(k_i+k_j)-i(\Oi+\Oj)}}
\nonumber\\
&&\times
\prod_{k<\ell} \left[1+e^{2\alpha+\beta(k_k+k_\ell)-i(\Ok+\Ol)}\right]
\nonumber\\
&=&\sum_{i<j}\bra \frac{e^{2\alpha+\beta(k_i+k_j)-i(\Oi+\Oj)}}{1+e^{2\alpha+\beta(k_i+k_j)-i(\Oi+\Oj)}}
\ket_{\bk,\bOmega}
\label{eq:L_kO}
\eea
where 
\be
\bra \cdot\ket_{\bk,\bOmega}=\frac{1}{Z} \sum_\bk \int_{-\pi}^\pi \dO \cdot \prod_{k<\ell} \left[1+e^{2\alpha+\beta(k_k+k_\ell)-i(\Ok+\Ol)}\right].
\ee
Hence, one has
\bea
\bra \cij\ket=\bra \frac{e^{2\alpha+\beta(k_i+k_j)-i(\Oi+\Oj)}}{1+e^{2\alpha+\beta(k_i+k_j)-i(\Oi+\Oj)}}
\ket_{\bk,\bOmega}
\label{eq:pij}
\eea
In the regime $\bra \cij\ket=\order{(N^{-1})}$, it is convenient to transform $\alpha\to \hat\alpha -\half \log (N/c)$, 
with $c=\order{(N^0)}$, to make (\ref{eq:pij}) explicitely $\order{(1/N)}$, 
so we get
\be
\bra c_{ij}\ket \simeq \frac{c}{N}\bra e^{2\hat\alpha+\beta(k_i+k_j)-i(\Oi+\Oj)}\ket_{\bk,\bOmega}
\ee
and
\bea
Z&=&\sum_\bk \int_{-\pi}^\pi \dO 
\exp\left[\frac{c}{N}\sum_{k<\ell} e^{2\hat\alpha+\beta(k_k+k_\ell)-i(\Ok+\Ol)}\right].
\label{eq:Z}
\eea
It is also useful to express the network distribution $p(\bc)$ in the scaled 
parameter $\hat\alpha$
\bea
p(\bc)&=&\frac{1}{Z}e^{2\alpha\sum_{i<j}\cij+\beta \sum_{i<j}\cij(k_i(\bc)+k_j(\bc))}
\nonumber\\
&=&\frac{1}{Z}\prod_{i<j}\left[
e^{2\alpha+\beta(k_i(\bc)+k_j(\bc))}\delta_{c_{ij},1}+\delta_{c_{ij},0}\right]
\nonumber\\
&=&\frac{1}{Z}\prod_{i<j}\left[
\frac{c}{N}e^{2\hat\alpha+\beta(k_i(\bc)+k_j(\bc))}\delta_{c_{ij},1}+\delta_{c_{ij},0}\right].
\label{eq:p_hat}
\eea
Next we introduce the following order parameters 
\begin{eqnarray}
P(k,\Omega|\bk,\bOmega)&=& \frac{1}{N}\sum_{r=1}^N \delta_{k,k_r}\delta(\Omega-\Omega_r)
\end{eqnarray}
and insert into (\ref{eq:Z}) for each $(k,\Omega)$ the following integral:
\begin{eqnarray}
1&=& \int\!\rmd P(k,\Omega)~\delta\Big[P(k,\Omega)\!-\!P(k,\Omega|\bk,\bOmega)\Big]
\nonumber
\\
&=&(N/2\pi)\int\!\rmd P(k,\Omega)\rmd\hat{P}(k,\Omega)\rme^{\rmi N\hat{P}(k,\Omega)P(k,\Omega)-
\rmi\sum_{r=1}^N \delta_{k,k_r}\delta(\Omega-\Omega_r)\hat{P}(k,\Omega)}
\end{eqnarray}
Discretizing $\Omega$ in steps of size $\Delta$ which is eventually sent to zero, we can then write the 
free-energy as the following path integral, with the short-hand $\{\rmd P\rmd\hat{P}\}=\prod_{k,\Omega}[\rmd P(k,\Omega)\rmd\hat{P}(k,\Omega)/2\pi]$ 
and sums over $\Omega$ transformed into integrals:
\bea
f&=&-\lim_{N\to \infty}\frac{1}{N} \log Z
\nonumber\\
&=&-\lim_{N\to\infty}\frac{1}{N} 
\log \intPP e^{iN\sum_{k\geq 0} \int_{-\pi}^\pi d\Omega P(k,\Omega)\hat P(k,\Omega)+ N \log \sum_{k\geq 0} \int d\Omega e^{i \Omega k - i \hPko}}
\nonumber\\
&&\times 
e^{\frac{Nc}{2}\sum_{k,k'\geq 0}\int_{-\pi}^\pi d\Omega d\Omega'\, \Pko P(k',\Omega')e^{2\hat\alpha+\beta(k+k')-i(\Omega+\Omega')}}
\nonumber\\
&=& -\lim_{N\to \infty}\frac{1}{N} \log \intPP e^{-N\Phi[P,\hP]}
\nonumber\\
\label{eq:f_int}
\eea
where 
\bea
\hspace*{-1cm}\Phi[P,\hP]&=&-i\sum_{k\geq 0} \int_{-\pi}^\pi d\Omega P(k,\Omega)\hat P(k,\Omega) -\frac{c}{2}\sum_{k,k'\geq 0}\int d\Omega d\Omega'\, \Pko P(k',\Omega')e^{2\hat\alpha+\beta(k+k')-i(\Omega+\Omega')}
\nonumber\\
\hspace*{-1cm}&&- \log \sum_{k\geq 0} \int_{-\pi}^\pi d\Omega e^{i \Omega k - i \hPko}
\eea
For large $N$, we can evaluate the path integral in (\ref{eq:f_int})
by steepest descent
\be
f={\rm min}_{P,\hat P}\Phi[P,\hat P]
\label{eq:extremization}
\ee 
Extremizing the action $\Phi$ over $P, \hat{P}$ leads to the saddle-point equations 
\bea
\Pko&=&\frac{e^{i \Omega k - i \hPko}}{\sum_{k\geq 0} \int_{-\pi}^\pi d\Omega e^{i \Omega k - i \hPko}}
\nonumber\\
-i \hPko&=&ce^{\hat\alpha +\beta k -i\Omega}\left(\sum_{k\geq 0} e^{\hat\alpha +\beta k}\int_{-\pi}^\pi d\Omega \Pko e^{-i \Omega} \right)
\eea 
The equation above suggests to define 
\be
P(k)=\int_{-\pi}^\pi d\Omega \Pko e^{-i \Omega}
\ee
and 
\be
\gamma(\hat\alpha,\beta)=\sum_{k\geq 0} e^{\hat\alpha +\beta k}P(k)
\ee
so that 
\be
-i \hPko=c e^{\hat\alpha +\beta k -i\Omega} \gamma(\hat\alpha,\beta)
\ee
Hence,
\bea
P(k,\Omega)=\frac{e^{i\Omega k +c\gamma e^{\hat\alpha +\beta k}e^{-i\Omega}}}{\sum_{k\geq 0} \int_{-\pi}^\pi d\Omega e^{i\Omega k +c\gamma e^{\hat\alpha +\beta k}e^{-i\Omega}}}
\eea
and 
\bea
P(k)=\frac{(c\gamma e^{\hat\alpha +\beta k})^{k-1}}{(k-1)!}\left[
\sum_{k\geq 0} \frac{(c\gamma e^{\hat\alpha +\beta k})^k}{k!}
\right]^{-1}\theta\left(k-\half\right)
\label{eq:Pk}
\eea
where $\theta(x)$ is the Heaviside 
step function, taking value $1$ for $x>0$ and $0$ for $x<0$.
Note that $P(k)$ is not a distribution because it is not normalised to one, instead we have $\sum_{k\geq 0}P(k)=\bra e^{\beta k}\ket_\gamma$ 
with
\be
\bra \cdot \ket_\gamma=\frac{\sum_{k\geq 0} \cdot (c\gamma e^{\hat\alpha+\beta k})^k/k!}{\sum_{k\geq 0} (c\gamma e^{\hat\alpha+\beta k})^k/k!}.
\ee
The resulting free energy density is
\be
f(\hat\alpha,\beta)=\frac{c\gamma^2(\hat\alpha,\beta)}{2}-\log \sum_{k\geq 0}\frac{[e^{\hat\alpha+\beta k}c\gamma(\hat\alpha,\beta)]^k}{k!}
\ee
where $\gamma$ solves the self-consistency equation
\be
c\gamma^2(\hat\alpha,\beta)=\bra k\ket_\gamma.
\label{eq:self_g}
\ee
Finally, the ensemble parameters $\hat\alpha,\beta$, have to be determined from the equations for the constraints which are found by   
taking the derivatives of $f$ with respect to $\alpha,\beta$ as
\bea
\kav&=&-\frac{\partial f}{\partial \alpha}=-\frac{\partial f}{\partial \hat\alpha}
=-c\gamma\frac{\partial\gamma}{\partial \hat\alpha}+\bra k\ket_\gamma+\frac{1}{\gamma}\bra k\ket_\gamma \frac{\partial\gamma}{\partial \hat\alpha}
=\bra k\ket_\gamma
\label{eq:partial_a}
\eea
and 
\bea
\bra k^2\ket&=&-\frac{\partial f}{\partial \beta}
=-c\gamma\frac{\partial\gamma}{\partial \beta}+\bra k^2\ket_\gamma+\frac{1}{\gamma}\bra k\ket_\gamma \frac{\partial\gamma}{\partial \beta}
=
\bra k^2\ket_\gamma
\label{eq:partial_b}
\eea
where the last equality in (\ref{eq:partial_a}) and (\ref{eq:partial_b}) follows from the saddle-point equation (\ref{eq:self_g}).
Combining (\ref{eq:partial_a}) and (\ref{eq:self_g}) we have $\gamma=\sqrt{\kav/c}$, yielding the below equations for the 
ensemble parameters: 
\bea
\kav=\frac{\sum_{k\geq 0} k (\sqrt{c\kav} e^{\hat\alpha+\beta k})^k/k!}{\sum_{k\geq 0} (\sqrt{c\kav} e^{\hat\alpha+\beta k})^k/k!}
\label{eq:kav_saddle_c}
\\
\bra k^2\ket=\frac{\sum_{k\geq 0} k^2 (\sqrt{c\kav} e^{\hat\alpha+\beta k})^k/k!}{\sum_{k\geq 0} (\sqrt{c\kav} e^{\hat\alpha+\beta k})^k/k!}
\label{eq:kvar_saddle_c}
\eea
Hence, we can finally write the free-energy density as 
\be
f(\kav,\kvar)=-\frac{c}{2}-\log \sum_{k\geq 0}e^{\alpha k +\beta k^2}g(k)
\label{eq:f_original}
\ee
where $\alpha$ and $\beta$ solve 
\bea
\kav=\frac{\sum_{k\geq 0} k g(k) e^{\alpha k+\beta k^2}}{\sum_{k\geq 0} g(k) e^{\alpha k +\beta k^2}}
\label{eq:kav_saddle}
\\
\bra k^2\ket=\frac{\sum_{k\geq 0} k^2 g(k) e^{\alpha k+\beta k^2}}{\sum_{k\geq 0} g(k) e^{\alpha k +\beta k^2}}
\label{eq:kvar_saddle}
\eea
with 
\be
g(k)=\frac{(\sqrt{c\kav})^k e^{-(\kav+c)/2}}{k!}.
\label{eq:g}
\ee
In conclusion, we have that for the finitely-connected two-star model, with constrained 
average connectivity $\kav$ and degree variance $\kvar$, 
the network distribution is
\bea
p(\bc|\kav,\kvar)&=&\frac{1}{Z}\prod_{i<j}\left[
\frac{c}{N}e^{2\alpha+\beta(k_i(\bc)+k_j(\bc))}\delta_{c_{ij},1}+\delta_{c_{ij},0}\right]
\label{eq:p_final}
\eea
where $\alpha$ and $\beta$ are determined from (\ref{eq:kav_saddle}) and 
(\ref{eq:kvar_saddle}).
We note that for the choice $c=\kav$, $g(k)$ becomes a Poissonian distribution with 
parameter $\kav$.

\subsection{Test for $\beta=0$}
First, we check the validity of our equations for the 
case $\beta=0$, where we should get back to the ER graphs. 
For $\beta=0$ the 
equation for the constraint (\ref{eq:kav_saddle}) gives 
\be
\bra k\ket =c e^{2\alpha}
\ee
Substituting in (\ref{eq:p_final}) we get
\bea
p(\bc)&=&\frac{1}{Z}\prod_{i<j}\left[
\frac{\kav}{N}\delta_{c_{ij},1}+\delta_{c_{ij},0}\right]
\eea
with 
$$Z=\sum_\bc \prod_{i<j}\left[
\frac{\kav}{N}\delta_{c_{ij},1}+\delta_{c_{ij},0}\right]=\prod_{i<j}\left[1+\frac{\kav}{N}\right]
$$
yielding, for large $N$, 
\bea
p(\bc)
&=&\prod_{i<j}\left[
\frac{\kav}{N}
\delta_{\cij,1}+\left(1-\frac{\kav}{N}\right)\delta_{\cij,0}
\right]
\eea
thus recovering the 
Erd\"os-R\'enyi ensemble.
%

\subsection{Numerical results}
\label{sec:numerics}
For $\beta\neq 0$, we need to solve equations (\ref{eq:kav_saddle}), (\ref{eq:kvar_saddle}) 
numerically. We note, however, that for positive values of $\beta$, the sums on the RHS of 
these equtions do not converge, hence only values $\beta\leq 0$ are admissible.
The parameter $c$ can be chosen arbitrarily, so we will 
set it to unit, without loss of generality. 

\begin{figure}
\centering
\includegraphics[width=180\unitlength]{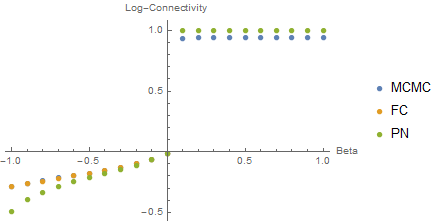}
\includegraphics[width=180\unitlength]{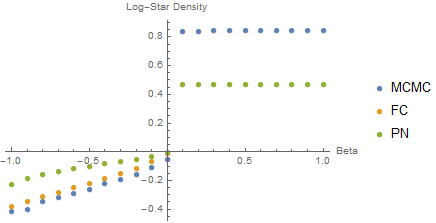}
\\
\includegraphics[width=180\unitlength]{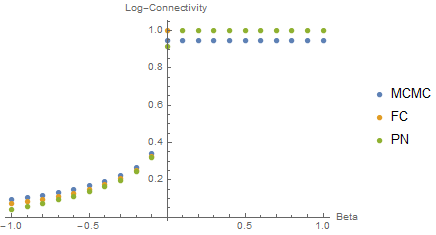}
\includegraphics[width=180\unitlength]{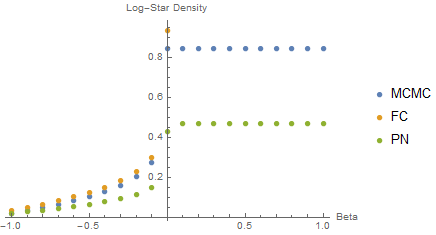}
\caption{Plot of $\log\kav/\log N$ (left panels) and $\log \kvar-\kav/\log N$ (right panels) as functions of the ensemble parameter 
$\beta$, for $\alpha=-0.5$ (top panels) and $\alpha=4$ (bottom panels) and $c=1$. MCMC denote Monte Carlo simulations for $N=3000$, FC denotes 
exact results from formulae (\ref{eq:kav_saddle}), (\ref{eq:kvar_saddle}) and PN denotes predictions from expansions about mean-field theory 
(\ref{eq:MF_kav}), (\ref{eq:MF_kvar}). 
}
\label{fig:MCMC}
\end{figure}

\begin{figure}
\centering
\includegraphics[width=160\unitlength]{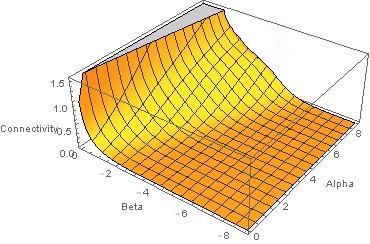}
\includegraphics[width=160\unitlength]{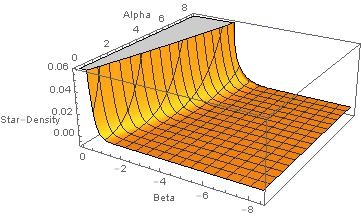}
\caption{Plot of the average connectivity (left panel) and the average 
density of stars (right panel) as functions of the ensemble parameters $\alpha, \beta$.
}
\label{fig:con_star}
\end{figure}

\begin{figure}
\centering
\includegraphics[width=240\unitlength]{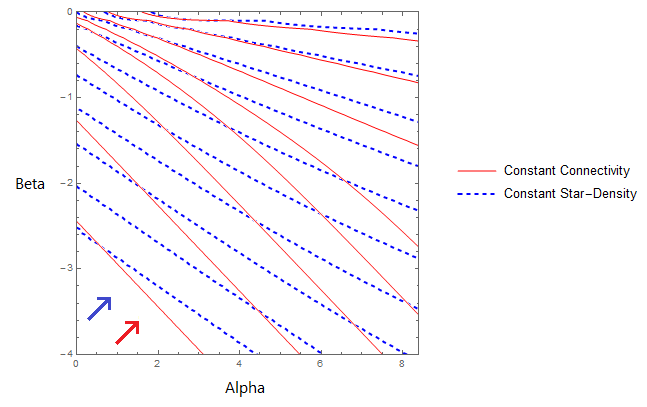}
\caption{Contour of constant average connectivity and average stars in the space of ensemble parameters $\alpha, \beta$.}
\label{fig:contours}
\end{figure}

In Figure \ref{fig:MCMC} we compare theoretical results from (\ref{eq:kav_saddle}), (\ref{eq:kvar_saddle}) (orange symbols) 
with MCMC simulations for networks of $N=3000$ nodes (blue symbols) and 
predictions from mean-field theory (\ref{eq:MF_kav}) and (\ref{eq:MF_kvar}) (green symbols).
Plots show the logarithm of the link and of the star densities normalised with the logarithm of the 
system size, as functions of $\beta$, for fixed values of $\alpha=-0.5, 4$, corresponding to 
low and high connectivity respectively.
MCMC simulations show a divergence in the links and stars densities for $\beta>0$, consistently with the fact that equations 
(\ref{eq:kav_saddle}), (\ref{eq:kvar_saddle}) do not converge in this regime, 
and show excellent agreement with theoretical predictions at 
$\beta\leq 0$. In particular, simulations data are on top 
of theoretical ones at low connectivity (top panel) and deviations 
stay within finite size effects at high connectivities (bottom panel).
In contrast, mean-field predictions are seen to perform well at high connectivity but, as expected, get very inaccurate for small connectivity.
In Figure \ref{fig:con_star}, we show three dimensional plots of the average connectivity and average density of stars, as functions of $\alpha$ and $\beta$.  
One has that for fixed values of $\alpha$, both connectivity and star density are at their highest for $\beta=0$ and they 
decay quickly for $\beta<0$. 
Notably, the star density is always finite while the connectivity is finite, hence the model fails to produce an arbitrary number of stars for any given finite connectivity. 
In particular, we find that the star density is always close to its physical minimum.
This is better understood by looking at contour plots of constant average connectivity and constant average star density in Figure \ref{fig:contours}. These show that for $\beta$ fixed the connectivity 
increases with $\alpha$ quicker than the star density, so that the star density decreases as we move along the contours of constant connectivity in the direction of decreasing $\beta$. 
Hence, the maximum number of stars is obtained for $\beta=0$, which 
corresponds to Erd\"os-R\'enyi graphs, satisfying $\bra k^2\ket-\kav=\kav^2$,  
so that the contour of constant connectivity and star density coincide.
The reason for this behavior
can be understood by looking at equation (\ref{eq:Pk}), showing that for 
$\beta \leq 0$,   
$P(k)$ is Poissonian ($\beta=0$) or narrower ($\beta<0$), whereas for $\beta>0$,
$P(k)$ is not normalizable, with an asymptotic behavior for large-$k$ given by 
$P(k)\sim e^{k(\beta k-\log k)}$, 
independent of $\alpha$. 

This shows that for $\kvar>\kav^2+\kav$, equations (\ref{eq:kav_saddle}) and (\ref{eq:kvar_saddle}) 
have no solution. We will see in the next section that in this range of the imposed constraints,
the partition function can no longer be calculated by the saddle-point method illustrated above and a 
different approach is needed. 
This leads to a phase diagram in the $\kav-\kvar$ plane with two phases 
which display different behaviors of the partition function 
$Z$ in the large $N$-limit. The critical line separating the two phases is 
found by solving (\ref{eq:kav_saddle}) and (\ref{eq:kvar_saddle}) for 
$\beta=0$, which gives $\kvar=\kav^2+\kav$. 
For $\kvar<\kav^2+\kav$
\be
Z(N\kav,N\kvar)\sim e^{-Nf(\kav,\kvar)}
\label{eq:Z_liquid}
\ee 
where $f(\kav,\kvar)$ is given by (\ref{eq:f_original}), whereas for $\kvar>\kav^2+\kav$ the asymptotic behavior 
of $Z$ is given by (\ref{eq:Z_cond}).
We will show in the next section that the 
phase transition occurring at $\kvar=\kav^2+\kav$ 
is a condensation from a liquid to a condensed phase, due to a large 
deviation of sums of random variables. 
We note that condensation may be avoided in the 
scaling regime $\beta=\order{(\log N/N)}$,  
the same where Strauss model is known to display non-trivial behavior \cite{Jonasson99}.

\section{Condensation as large deviation of sums of random degrees}
\label{sec:condensation}
In this section we show that the condensation transition 
occurring at $\kvar=\kav^2+\kav$ is related to a large deviation 
of sums of random degrees and we provide the asymptotics of 
$Z$ in the condensate phase $\kvar>\kav^2+\kav$. 
To this purpose, it is convenient to use an alternative but equivalent definition of the 
two-star model, where the constraints on the links and stars are implemented directly in the network distribution
\bea
\hspace*{-2.5cm}p(\bc|\kav,\kvar)=\frac{1}{Z(N\kav,N\kvar)}\prod_{i<j}\left[\frac{c}{N}\delta_{c_{ij},1}+\left(1-\frac{c}{N}\right)\right]\delta(\sum_i k_i(\bc)-N\kav)\delta(\sum_i k_i^2(\bc)-N\kvar)
\nonumber\\
\label{eq:alt_p}
\eea
via Khronecher deltas and links are drawn otherwise randomly and independently with likelihood $c/N$.
By using path integrals and the saddle point method used in Section \ref{sec:exact} we find that in the fluid phase $\kvar<\kav^2+\kav$
the partition function $Z(N\kav,N\kvar)$ has the large deviation behavior 
(\ref{eq:Z_liquid})
with rate function given by the free-energy density (see 
\ref{app:alternative} for full details)
\be
f(\kav,\bra k^2\ket)=\alpha \kav+\beta \kvar -\log \sum_k e^{\alpha k+\beta k^2}g(k)
\label{eq:f_alt}
\ee
where $\alpha, \beta$ are determined from equations (\ref{eq:kav_saddle}), (\ref{eq:kvar_saddle}) and $g(k)$ is defined in (\ref{eq:g}).
Up to additive constants, the free-energy density (\ref{eq:f_alt}) 
is identical to (\ref{eq:f_original}), 
hence the definition (\ref{eq:alt_p}) of the two-star model is
thermodynamically equivalent to the standard definition (\ref{eq:p_final}).
The marginal distribution $p(k)$ in the fluid phase is given by 
(see \ref{app:marginal} for derivations)
\be
p(k)=g(k) \frac{e^{\alpha k +\beta k^2}}{\sum_{k\geq 0} g(k) e^{\alpha k +\beta k^2}}
\label{eq:marg}
\ee
showing that all random degrees 
contribute to the sums $\sum_i k_i$, $\sum_i k_i^2$ with 
small values.

Notably, the free-energy (\ref{eq:f_alt}) and the marginal distribution 
(\ref{eq:marg}) are identical to those of the system 
of independently and identically distributed random variables 
$\bk=\{k_1,\ldots,k_N\}$
with "bare" distribution $\hat g(k)=g(k)/\sum_k g(k)$ 
and global constraints on the 
average and the variance, introduced in \cite{Majumdar14} 
\be
\hspace*{-1cm}p(\bk|\kav,\kvar)=\frac{1}{Z(N\kav,N\kvar)}\prod_i \hat g(k_i)\delta(\sum_i k_i-N\kav)\delta(\sum_i k_i^2-N\kvar)
\label{eq:factorised}
\ee
The connection between the two models is transparent: a graph drawn 
from (\ref{eq:alt_p}) will have a degree sequence $\bk$ where the 
degrees are random variables drawn indipendently from the Poissonian 
distribution $\hat g(k)$ with average $\sqrt{c\kav}$, subject to 
the two global constraints. For the 
choice $c=\kav$ each degree will be a Poissonian variable with average 
$\kav$, subject to constraints.
  
Systems with factorised steady states (\ref{eq:factorised}) 
are known to display condensation 
for non-heavy-tailed distributions $\hat g(k)\sim e^{-rk^\gamma}$ with
$1\leq \gamma <2$, 
when conditioned on large deviations of their linear statistics, i.e. for 
$\kvar$ larger than a critical value $\sigma(\kav)$, which is 
found by solving 
(\ref{eq:kav_saddle}) and (\ref{eq:kvar_saddle}) for $\beta=0$ 
\cite{Majumdar14}, i.e. from
\be
\sigma(\kav)=\frac{\sum_k k^2 g(k)e^{\alpha k}}{\sum_k g(k)e^{\alpha k}}
\label{eq:sigma}
\ee
with $\alpha$ solving
\be
\kav=\frac{\sum_k k g(k)e^{\alpha k}}{\sum_k g(k) e^{\alpha k}}.
\label{eq:kavc}
\ee
For $g(k)$ defined in (\ref{eq:g}), the critical value $\sigma(\kav)$ can be calculated 
exactly: equation (\ref{eq:kavc}) gives 
$\alpha=\ln\sqrt{\kav/c}$, and substituting in (\ref{eq:sigma}) we obtain 
$\sigma(\kav)=\kav^2+\kav$.
 
For $\kvar>\sigma(\kav)$, the method developed in \cite{Majumdar14} yields for 
function (\ref{eq:g}) with asymptotics $g(k)\sim e^{-k\log k}$, the following asymptotic 
behaviour for the partition function 
\be
Z(N\kav,N\kvar)\sim e^{-N I(\kav)-
[N\kvar-N\sigma(\kav)]^\half \log [N\kvar-N\sigma(\kav)]^\half} 
\label{eq:Z_cond}
\ee
with 
$$
I(\kav)=\alpha \kav -\log \sum_{k\geq 0} \hat g(k) e^{\alpha k}
$$
and $\alpha$ determined from (\ref{eq:kavc}).
In this regime, the marginal 
$p(k)$ shows a bump at $k\sim \sqrt{N}$, 
meaning that there is a condensate 
of size $\sim \sqrt{N}$ residing on a single site. 
Sampling networks from the two-star distribution (\ref{eq:alt_p}) 
will then lead to graphs which have a bulk of homogeneous degrees 
and one single site accounting for all the degree hetoregeneity in the 
network, making this model unsuitable as a null model for finitely 
connected random graphs with constrained average degree and variance. 
In addition, although Markov procsses generating random variables $\bk$ 
under the two global constraints can be constructed as in \cite{Majumdar14},
the definition of algorithms to sample networks $\bc$
from the measure (\ref{eq:alt_p}) with the two hard constraints on degree average and variance, 
in an efficient and unbiased way, poses a challenge in 
its own right.
We note that removing the hard constraint on the variance we obtain the 
system with factorised states studied in \cite{Zia06}
\be
p(\bk|\kav)=\frac{1}{Z(N\kav)}\prod_i \hat g(k_i)\delta(\sum_i k_i(\bc)-N\kav)
\ee
which is known to exhibit 
condensation only for 
heavy-tailed bare distributions $\hat g(k)\sim A k^{-\gamma}$, with $\gamma>2$, 
for $\kav<\sum_{k\geq 0} k \hat g(k)$.
For these models, the "dressed" marginal distribution is found to be 
\cite{Zia06}
\be 
p(k)=\frac{\hat g(k)e^{\alpha k}}{\sum \hat g(k)e^{\alpha k}}
\ee 
with $\alpha$ solving $\kav=\sum_k k p(k)$.
This suggests that condensation transitions in the two-star model may be 
avoided 
by removing the hard constraint on the variance and choosing the bare 
distribution 
$\hat g(k)$ in such a way that the "dressed" marginal 
displays the desired variance $\kvar=\sum_k p(k)k^2$ while satisfying 
$\sum_{k\geq 0}k \hat g(k)<\kav$.

\section{Conclusion}
\label{sec:conclusions}
In this work we analysed the finitely connected $2$-star model. This model has been solved analytically within mean-field approximations, 
which predict a second-order phase transition between a symmetric and a symmetry-broken phase and a 
first-order transition in the link density occuring along the critical line separating the symmetry-broken phases, in the ensemble parameter space. 
In this work we solved the $2$-star model exactly, in the thermodynamic limit, in the finite connectivity regime, where mean-field approximations 
are shown to become inexact. 
Our results show that in the thermodynamic limit the system undergoes a condensation transition, from a liquid to a condensed phase, 
related to large deviations in the sum of the random degrees, 
induced by the global constraints on their linear statistics. 
We showed that in the liquid phase, the degree statistics exhibited 
by the model is always in between regular and Erd\"os-R\'enyi graphs.
Our results are in excellent agreement with MCMC simulations and are 
compared with existing results from mean-field theory and expansions around it. 
The latter 
become inaccurate for small connectivity, albeit providing the correct 
location of the critical line.
When the global constraints imposed on the linear statistics 
insist on a degree statistics 
more heterogeneous than Erd\"os-R\'enyi graphs, the model undergoes a 
condensation transition, whereby a condensate 
residing on a single site appears 
in the system, 
which accounts for all the degree heterogeneity of the network, while 
the bulk of the network have homogeneous degrees.

We conclude that the finitely connected $2$-star model is 
unsuitable, in its standard definition, as a null-model for real networks,
with prescribed connectivity and link density. We suggest 
possible modifications of the model which may lead to non-trivial degree 
statistics in the finite connectivity regime, which 
either entail a different scaling of the 
Lagrange multiplier $\beta$ constraining the stars or the use of 
a soft constraint for the star density, while the 
link density is hardly constrained. 

\section{Aknowledgments}
AA wishes to thank Peter Sollich for useful discussions.

\section{References}
\bibliographystyle{plain}
\bibliography{GraphGeneration}

\appendix

\section{Mean-field predictions in the finitely connected regime}
\label{app:graphical}
In the low connectivity regime we demand that $p=c/N$ with 
$c=\order{(N^0)}$. 
To ensure that, we see from (\ref{eq:PN}) that we have to set $\beta=\order{(1)}$ and scale the 
ensemble parameter $\alpha$ as 
$\alpha\to \hat\alpha -1/2 \log N$ so that the RHS of (\ref{eq:PN}) becomes
\bea
\frac{1}{2}\Big[\tanh\big(\beta c + \hat\alpha -\half 
\log N\big) + 1\Big]&=&\frac{1}{2}\left[\frac{\frac{1}{N}e^{2\beta c + 
2\hat\alpha}-1}{\frac{1}{N}e^{2\beta c + 2\hat\alpha}+1} + 1\right]
=\frac{\frac{1}{N}e^{2\beta c + 2\hat\alpha}}{\frac{1}{N}e^{2\beta c + 2\hat\alpha}+1} 
\nonumber\\
&\sim& \frac{1}{N}e^{2\beta c + 2\hat\alpha}
\eea
where the last equality holds for large $N$ and yields $p=\order{(N^{-1})}$ as required.
Equating this to the LHS of (\ref{eq:PN}) we get the self-consistency equation
\begin{equation}
c=e^{2\beta c + 2\hat\alpha}
\label{eq:PN2}
\end{equation}

A graphical analysis of (\ref{eq:PN2}) reveals that 
for 
$\beta\leq 0$ there is only one solution, whereas for $\beta>0$ there can be one, two or no solutions depending on the choice of 
$\hat\alpha$ and $\beta$. To identify the regions of the parameters for which solutions exist, note that the RHS of (\ref{eq:PN2}) is convex and strictly increasing in $c$. 
Let $c^\star$ be the value of $c$ for which 
$\frac{\partial}{\partial c}RHS= 2\beta e^{2\beta c + 2\hat\alpha}=1$. 
If the value of the RHS at $c^\star$ is greater than $c^\star$ 
then there can be no solutions of the self-consistency equation, 
while if it is less than $c^\star$ there are two, and there is only one 
solution if the RHS equals $c^\star$. 
We find that $c^\star$ is given by
\be
2\beta e^{2\beta c^\star+2\hat\alpha}=1
\label{eq:cstar}
\ee
We have that there exists only one solution when 
$c^\star=e^{2\beta c^\star + 2\hat\alpha}$,
which using the defining equation of $c^\star$ (\ref{eq:cstar}) simplifies to  
$c^\star=1/(2\beta)$.
Inserting in (\ref{eq:cstar}) we find that there is a unique solution for 
$\beta=\half e^{-1-2\hat\alpha}$.
We have no solutions when
$
c^\star<e^{2\beta c^\star + 2\hat\alpha}=1/(2\beta)
$
which inserted in (\ref{eq:cstar}) gives 
$1<2\beta e^{1+2\hat\alpha}$
hence 
$
\beta > \half e^{-1-2\hat\alpha}$.
Finally we have two solutions for $c^\star>e^{2\beta c^\star + 2\hat\alpha}$, that is in the region 
$\beta < \half e^{-1-2\hat\alpha}$.
This implies that at $\beta=e^{-1-2\hat\alpha}/2$ the only solutions of 
(\ref{eq:PN2}) is $c=e^{1+2\hat\alpha}$, whereas for $\beta< e^{-1-2\hat\alpha}/2$ 
there are two solutions, one smaller and one greater  
than $e^{1+2\hat\alpha}$. As $\beta$ approaches zero the greater solution tends to 
infinity while the lower solution tends to $e^{2\hat\alpha}$. 
From (\ref{eq:MF_p}) we have $e^\lambda=1/(1-p)$ which substituted in (\ref{eq:MF_f}) gives 
for the free-energy density $f=F/N=\ln(1-p) (N-1)/2 \simeq -c/2$, for $p=c/N$ and $N\gg c$, 
showing that the free-energy decreases as the connectivity increases. Hence, the stable solution 
for $0\leq \beta \leq e^{-1-2\hat\alpha}/2$ will be the one with higher connectivity.
In conclusion, the mean-field theory in the finite connectivity regime predicts a critical line at $\beta=0$ where the average connectivity jumps from $\order{(1)}$ to $\order{(N)}$ values.
Although the mean-field approximation is invalid in the finite connectivity regime, it picks up the correct location of the critical line $\beta=0$ 
found from the exact analysis. 

\section{Calculation of the partition function in the liquid 
phase}
\label{app:alternative}
In this section we calculate the normalising constant $Z(N\kav,N\kvar)$ 
of the distribution 
(\ref{eq:alt_p}). First off, we use the identity $1=\sum_\bk \delta_{\bk,\bk(\bc)}$ to write
\bea
\hspace*{-2.5cm}
Z(\bc|N\kav,N\kvar)=\sum_\bk \sum_\bc \delta_{\bk,\bk(\bc)}\prod_{i<j}\left[\frac{c}{N}\delta_{c_{ij},1}+\left(1-\frac{c}{N}\right)\right]\delta(\sum_i k_i(\bc)-N\kav)\delta(\sum_i k_i^2(\bc)-N\kvar)
\nonumber\\
\eea
and the Fourier representation of the Kronecher deltas, giving 
\bea
Z(\bc|N\kav,N\kvar)&=&\int_{-\pi}^\pi \frac{d\omega d\omega'}{4\pi^2} e^{i\omega N \kav+i\omega' N \kvar}\sum_\bk \int_{-\pi}^\pi \dO e^{-i\omega \sum_i k_i -i\omega' \sum_i k_i^2}
\nonumber\\
&&\times
\sum_\bc \prod_{i<j}\left[\frac{c}{N}e^{-i(\Omega_i+\Omega_j)}\delta_{c_{ij},1}+\left(1-\frac{c}{N}\right)\right]
\label{eq:alt_Z}
\eea
We next introduce the following order parameters 
\begin{eqnarray}
P(\Omega|\bOmega)&=& \frac{1}{N}\sum_{r=1}^N \delta(\Omega-\Omega_r)
\label{eq:parameter}
\end{eqnarray}
and insert into (\ref{eq:alt_Z}) for each $\Omega$ the following integral:
\begin{eqnarray}
1&=& \int\!\rmd P(\Omega)~\delta\Big[P(\Omega)\!-\!P(\Omega|\bOmega)\Big]
\nonumber
\\
&=&(N/2\pi)\int\!\rmd P(\Omega)\rmd\hat{P}(\Omega)\rme^{\rmi N\hat{P}(\Omega)P(\Omega)-
\rmi\sum_{r=1}^N \delta(\Omega-\Omega_r)}
\end{eqnarray}
Discretizing $\Omega$ in steps of size $\Delta$ which is eventually sent to 
zero, and proceeding as in Sec. \ref{sec:exact}, we can then write the 
free-energy density as the path integral
\bea
f&=&
-\lim_{N\to\infty}\frac{1}{N} \log \int_{-\pi}^\pi \frac{d\omega d\omega'}{4\pi^2} e^{i\omega N \kav+i\omega' N \kvar}
\nonumber\\
&&\times \intPP e^{iN \int d\Omega P(\Omega)\hat P(\Omega) +\frac{cN}{2}\left[\int d\Omega d\Omega'\, P(\Omega) P(\Omega')e^{-i(\Omega+\Omega')}-1\right]}
\nonumber\\
&&\times \prod_i \sum_{k_i}\int \frac{d\Omega_i}{2\pi}e^{i\Omega_i k_i  - i\omega k_i -i\omega' k_i^2  -i \hat P(\Omega)}
\nonumber\\
&=&-\lim_{N\to\infty}\frac{1}{N} 
\log \int_{-\pi}^\pi \frac{d\omega d\omega'}{4\pi^2}\intPP 
e^{-N\Phi(P,\hP,\omega,\omega')}
\eea
where 
\bea
\hspace*{-1cm}\Phi(P,\hP,\omega,\omega')&=&-i\omega \kav-i\omega' \kvar -i\int_{-\pi}^\pi d\Omega P(\Omega)\hat P(\Omega) -\frac{c}{2}\int d\Omega d\Omega'\, P(\Omega) P(\Omega')e^{-i(\Omega+\Omega')}
\nonumber\\
&&- \log \sum_{k\geq 0} \int_{-\pi}^\pi d\Omega e^{i \Omega k - i\omega k -i\omega' k^2 -i \hat P(\Omega)}+\frac{c}{2}
\label{eq:Phi}
\eea
For large $N$, we can evaluate the integral by steepest descent
\be
f={\rm min}_{P,\hat P,\omega,\omega'}\Phi(P,\hat P,\omega,\omega')
\ee 
Extremizing the action $\Phi$ over $P, \hat{P}$ we obtain, proceding as in Section \ref{sec:exact}, 
\be
c\gamma^2=\bra k \ket_\gamma
\ee
with
\be
\bra \cdot \ket_\gamma=\frac{\sum_{k\geq 0} \cdot (c\gamma)^k e^{-i\omega k-i\omega' k^2}/k!}{\sum_{k\geq 0} (c\gamma)^k e^{-i\omega k-i\omega' k^2}/k!}.
\ee
Extremizing $\Phi$ over $\omega, \omega'$ we obtain
\bea
\kav=\bra k \ket_\gamma,\quad\quad\quad \bra k^2\ket=\bra k^2 \ket_\gamma
\eea
Substituting the saddle point equations in the free energy we have 
\bea
f&=&-i\omega \kav-i\omega' \kvar 
-\log \sum_{k\geq 0} \int_{-\pi}^\pi d\Omega e^{- i\omega k -i\omega' k^2}g(k)
\eea
with 
\be
g(k)=\frac{(\sqrt{c\kav})^k e^{-(c+\kav)/2}}{k!}
\ee
and $\omega,\omega'$ determined from 
\bea
\kav=\frac{\sum_{k\geq 0} k g(k) e^{-i\omega k-i\omega' k^2}}{\sum_{k\geq 0} g(k) e^{-i\omega k -i\omega'  k^2}}
\label{eq:alt_kav}
\\
\bra k^2\ket=\frac{\sum_{k\geq 0} k^2 g(k) e^{-i\omega k-i\omega' k^2}}{\sum_{k\geq 0} g(k) e^{-i\omega k -i\omega' k^2}}
\label{eq:alt_kvar}
\eea
Setting $-i\omega=\alpha$, $-i\omega'=\beta$ yields (\ref{eq:f_alt}).

\section{Calculation of the marginal distribution in the liquid phase}
\label{app:marginal}
In this section we calculate the marginal distribution
\be
p(k|\bc)=\frac{1}{N}\sum_i \delta_{k,k_i(\bc)}
\ee
In the limit of large $N$, we expect this quantity to converge 
to its ensemble average 
\bea
p(k)&=&\bra p(k|\bc)\ket=\int \frac{dx}{2\pi} e^{ixk}\int_{-\pi}^\pi \frac{d\omega d\omega'}{4\pi^2} e^{i\omega N \kav+i\omega' N \kvar}
\nonumber\\
&&\times \sum_\bk \int_{-\pi}^\pi \dO e^{-i\omega \sum_i k_i -i\omega' \sum_i k_i^2}
\nonumber\\
&&\times
\sum_\bc \prod_{k<\ell}\left[\frac{c}{N}e^{-i(\Omega_k+\Omega_\ell)-ix(\delta_{i\ell}+\delta_{ik})}\delta_{c_{k\ell},1}+\left(1-\frac{c}{N}\right)\delta_{c_{k\ell},0}\right]
\nonumber\\
&=&\int \frac{dx}{2\pi} e^{ixk}\int_{-\pi}^\pi \frac{d\omega d\omega'}{4\pi^2} e^{i\omega N \kav+i\omega' N \kvar}
\nonumber\\
&&\times \sum_\bk \int_{-\pi}^\pi \dO e^{-i\omega \sum_i k_i -i\omega' \sum_i k_i^2}
\nonumber\\
&&\times
\exp\left\{\sum_{k<\ell}\frac{c}{N}\left[e^{-i(\Omega_k+\Omega_\ell)-ix(\delta_{i\ell}+\delta_{ik})}-1\right]\right\}
\nonumber\\
\eea
Next we work out the curly brackets as 
\bea
\sum_{k<\ell}\frac{c}{N}\left[e^{-i(\Omega_k+\Omega_\ell)-ix(\delta_{i\ell}+\delta_{ik})}-1\right]=
\sum_{k<\ell}\frac{c}{N}\left\{e^{-i(\Omega_k+\Omega_\ell)}[(e^{-ix}-1)(\delta_{i\ell}+\delta_{ik})+1]-1\right\}
\nonumber\\
\eea
Inserting the order parameter (\ref{eq:parameter}) via path integrals as done in \ref{app:alternative}
we arrive at 
\bea
\hspace*{-1cm}
p(k)&=&\int \frac{dx}{2\pi} e^{ixk}\int_{-\pi}^\pi \frac{d\omega d\omega'}{4\pi^2} e^{i\omega N \kav+i\omega' N \kvar}
\intPP e^{iN \int d\Omega P(\Omega)\hat P(\Omega) +\frac{cN}{2}\left[\int d\Omega d\Omega'\, P(\Omega) P(\Omega')e^{-i(\Omega+\Omega')}-1\right]}
\nonumber\\
&&\times \frac{1}{N} \sum_i  \sum_{\bk}\int \dO e^{- i\omega \sum_i k_i -i\omega' \sum_i k_i^2  -i \sum_i \hat P(\Omega_i)+
ce^{-i\Omega_i}(e^{-ix}-1)\int d\Omega P(\Omega)e^{-i\Omega}}
\nonumber\\
&=&
\int \frac{dx}{2\pi} e^{ixk}\int_{-\pi}^\pi \frac{d\omega d\omega'}{4\pi^2} e
^{i\omega N \kav+i\omega' N \kvar}
\intPP e^{iN \int d\Omega P(\Omega)\hat P(\Omega) +\frac{cN}{2}\left[\int d\Omega d\Omega'\, P(\Omega) P(\Omega')e^{-i(\Omega+\Omega')}-1\right]}
\nonumber\\
&&\times \frac{1}{N} \sum_i  
\frac{\sum_{k_i}\int d\Omega_i e^{-i\Omega_i k_i - i\omega k_i -i\omega' k_i^2  -i \hat P(\Omega_i)+ce^{-i\Omega_i}(e^{-ix}-1)\int d\Omega P(\Omega)e^{-i\Omega}}}
{\sum_{k_i}\int d\Omega_i e^{-i\Omega_i k_i- i\omega k_i -i\omega' k_i^2  -i \hat P(\Omega_i)}}
\nonumber\\
&&\times\prod_j
\sum_{k_j}\int d\Omega_j e^{-i\Omega_j k_j- i\omega k_j -i\omega' \sum k_j^2  
-i \hat P(\Omega_j)}
\nonumber\\
&=&
\int \frac{dx}{2\pi} e^{ixk}\int_{-\pi}^\pi \frac{d\omega d\omega'}{4\pi^2} 
\intPP e^{-N\Phi([P,\hP],\omega,\omega')} 
\nonumber\\
&&
\times\frac{\sum_k\int d\Omega e^{i\Omega k- i\omega k -i\omega' k^2  -i \hat P(\Omega)+ce^{-i\Omega}(e^{-ix}-1)\int d\Omega P(\Omega)e^{-i\Omega}}}{\sum_k\int d\Omega e^{i\Omega k- i\omega k -i\omega' k^2  -i \hat P(\Omega)}}
\label{eq:marginal}
\eea
where $\Phi(P,\hP,\omega,\omega')$ is as in (\ref{eq:Phi}).
We next calculate the integrals over $\omega,\omega',P,\hat P$ by steepest 
descent.
At the saddle point we have
\be
-i \hP(\Omega)= e^{-i\Omega} \sqrt{c\kav}
\ee
and we obtain 
\bea
p(k)&=&\int \frac{dx}{2\pi} e^{ixk}
\frac{\sum_q\int d\Omega e^{i\Omega q- i\omega q -i\omega' q^2 +\sqrt{c\kav}e^{-i\Omega}e^{-ix}}}{\sum_q\int d\Omega e^{i\Omega q- i\omega q -i\omega' q^2  +
\sqrt{c\kav} e^{-i\Omega}}}
\label{eq:marginal}
\eea
where $\omega,\omega'$ solve (\ref{eq:alt_kav}), (\ref{eq:alt_kvar}). 
Performing the $\Omega$-integrals we have 
\bea
p(k)&=&\int \frac{dx}{2\pi} e^{ixk}
\frac{\sum_q e^{-i\omega q -i\omega' q^2}  (ce^{-ix}\gamma)^q/q!}{\sum_q  e^{- i\omega q -i\omega' q^2} (c\gamma)^q/q!}
\label{eq:marginal}
\eea
and carrying out the integration over $x$ finally gives 
\bea
p(k)&=&\frac{e^{-i\omega k -i\omega' k^2} (\sqrt{c\kav})^k/k!}{\sum_q e^{- i\omega q -i\omega' q^2} (\sqrt{c\kav})^q/q!}
\nonumber\\
&=& g(k) \frac{e^{-i\omega k -i\omega' k^2}}{\sum_k g(k)e^{-i\omega k -i\omega' k^2}}
\eea
with $g(k)$ defined in (\ref{eq:g}). Setting $-i\omega=\alpha$, $-i\omega'=\beta$ finally yields (\ref{eq:marg}).

\end{document}

%% file: commands.tex
\newcommand{\be}{\begin{equation}}
\newcommand{\ee}{\end{equation}}
\newcommand{\bd}{\begin{displaymath}}
\newcommand{\ed}{\end{displaymath}}
\newcommand{\bea}{\begin{eqnarray}}
\newcommand{\eea}{\end{eqnarray}}
\newcommand{\bean}{\begin{eqnarray*}}
\newcommand{\eean}{\end{eqnarray*}}

\newcommand{\bra}{\langle}
\newcommand{\ket}{\rangle}

\newcommand{\order}{{\mathcal O}}

\newcommand{\kav}{\langle k\rangle}

\newcommand{\bc}{{\bf c}}

\newcommand{\bk}{{\bf k}}

\newcommand{\blambda}{\mbox{\protect\boldmath $\lambda$}}

\newcommand{\bOmega}{\mbox{\protect\boldmath $\Omega$}}

\newcommand{\half}{{\frac{1}{2}}}
\newcommand{\cij}{{c_{ij}}}
\newcommand{\cji}{{c_{ji}}}
\newcommand{\cjk}{{c_{jk}}}

\newcommand{\dO}{{\frac{d\bOmega}{(2\pi)^N}\rme^{i\bOmega \bk}}}
\newcommand{\Oi}{{\Omega_i}}
\newcommand{\Oj}{{\Omega_j}}
\newcommand{\Ok}{{\Omega_k}}
\newcommand{\Ol}{{\Omega_\ell}}
\newcommand{\hP}{{{\hat P}}}
\newcommand{\hPko}{{{\hat P}(k,\Omega)}}
\newcommand{\Pko}{{P(k,\Omega)}}

\newcommand{\intPP}{{\int \{dP d\hP\}}}
\newcommand{\bark}{{\bar k}}